\documentclass[preprint2]{aastex}

\begin{document}

\title{Corotation: its influence on the chemical abundance pattern of the Galaxy}
\author{Yu.N.Mishurov\footnote{ Space Research Department, Rostov State University,
 5 Zorge, Rostov-on-Don, 344090, Russia; E-mail: mishurov@phys.rnd.runnet.ru}, 
 J.R.D.L\'epine\footnote{ Instituto de Astronomia, Geof\'isica e Ci\^encias Atmosf\'ericas, Universidade
 de São Paulo, Cidade Universit\'aria, S\~ao Paulo, SP, Brazil; E-mail: jacques@iagusp.usp.br}
\and
I.A.Acharova\footnote{Space Research Department, Rostov State University, 5 Zorge, Rostov-on-Don,
 344090, Russia; E-mail: kfk@phys.rnd.runnet.ru}}

\begin {abstract} 
A simple theory for the chemical enrichment of the Galaxy which takes into account the 
effects of spiral arms on heavy elements output was developed. In the framework of the
model with the corotation close to the position of the Sun in the Galaxy the observed abundance
features are explained.
\end{abstract}



\section{Introduction} 

The radial distribution of heavy elements in the Galaxy is a  signature of the Galactic history and structure.
Especially important information is brought by various features in the abundance pattern.

In early papers the radial variation of abundance was approximated by a function close to the linear one
in logarithmic scale (see e.g. Harris, 1981; Harris \& Pilachowski, 1984 and references therein). However,
in the last decade some features of the chemical distribution were revealed that do not support the simple
linear model. Indeed, Simpson et al (1995) and V\'ilchez \& Esteban (1996) derived a plateau-like (or step-like)
abundance structure of the H II regions. The same pattern was found by Twarog et al (1997) for open clusters,
by Smart \& Rollestone (1997) for B stars, by Caputo et al. (2001) for Cepheids. In recent papers Andrievsky
et al. (2002, a,b) on the basis of precise spectroscopic studies of Cepheids derived a "bimodal" abundance
gradient with a plateau beyond the solar distance and a sufficiently steep gradient inside the solar circle.
Finally, Maciel \& Quireza (2000) found a parabolic metallicity distribution for the planetary nebulae,
with a minimum near the solar distance.

What is the mechanism of such features formation? In this paper we will inspect the idea that the bimodal
abundance gradient can be a consequence of the corotation resonance.The underlying idea is that at corotation
the star formation rate presents a minimum, and consequently, the metallicity accumulated after a few billion
years at this radius can also have minimum. Bearing in mind that since Marochnik et al (1972) and  Cr\'ez\'e \& 
Mennessier (1973) a lot of studies showed that the corotation is situated very close to the Sun 
(Nelson \& Matsuda, 1976; Mishurov et al, 1979; Mishurov et al, 1997, Amaral \& L\'epine, 1997; von Linden
et al., 1998; Leicht \& Vasisht, 1998; Mishurov \& Zenina, 1999a,b; L\'epine et al, 2001; Fernandez et al.,
2001 etc) it would be interesting to analyze the influence of the corotation on the abundance pattern. 

On the other hand, according to Weinberg (1994), Englmaier \& Gerhard (1999), Dehnen (2000) and others the
corotation is situated very close to the galactic center at a distance $\sim$ 4 kpc. So, investigating 
the chemical abundance structure of the Galactic disk, we can hope to make a choice between the above models. 

\section{Basic idea and equations}

Standard equations governing the chemical evolution of the galactic disk take into account both the creation
of heavy elements and mass transfer between gaseous and stellar components (Tinsley, 1980). They are 
essentially the equations of continuity and convey the mass conservation law for a mixture: gas + stars 
+ heavy elements (e.g. Lacey \& Fall, 1989, hereafter LF; Portinari \& Chiosi, 1999, hereafter PC). By means
of these equations researchers usually try to simultaneously explain both the radial behavior of metal
abundance and gaseous and stellar distributions. However the corresponding equations must be completed by the
dynamical ones for instance by the Euler gasdynamic equation, the Boltzman equation for the stellar component,
the Poisson equation for gravitation field etc. Only such an approach would enable us to derive all radial profiles
in a self-consistent manner. Of course, this way is very complicated and most authors prefer to exclude
the dynamical part of the problem (a rare exception is the paper of Martinet \& Friedly, 1997) and consider
the task in the framework of the standard scheme, specifying some model laws for the star formation rate (SFR),
the infall rate of matter onto the Galactic disk, and the radial flows within the disk, so as to explain 
the observed radial distribution for chemical abundance, star density and interstellar gas density.

To make more prominent the effects of the spiral arms and of the corotation on the abundance pattern which
are not masked by other factors, in the present paper we shall only concentrate our attention on the chemical
enrichment. For this task we adopt the stationary model of Haywood et al. (1997) for Galactic evolution.
According to these authors, gas density and infall rate are constant in time. This enables us to consider 
the chemical part of the problem separately for various laws of infall rate, radial flow and star formation
rate (SFR) \footnote {Of course, we abstain from explaining the stellar density distribution. In any case
it is impossible to adequately derive the complicated structure of the stellar disk in the framework of
models which do not take into account dynamical processes like Galactic contraction as a whole, influence
of the bar and of the spiral structure etc.}. The corresponding equation we write in a form close to that
of LF:

\begin{eqnarray}
\mu_g \frac{\partial Z}{\partial t}  & = &  P_Z\psi + f(Z_f-Z) - \mu_g u \frac{\partial Z}{\partial R} + 
\nonumber\\
& & \frac{1}{R} \frac {\partial}{\partial R}(\mu_g D\frac{\partial Z}{R}),
\end{eqnarray}
where $\mu_g$ is the surface density of interstellar gas, $Z$ is the abundance of heavy elements,
$\psi$ is the star formation rate, $P_Z$ is the mass fraction of a star ejected as newly synthesized heavy elements (Tinsley, 1980), $u$ is the macroscopic radial
velocity of the gas, $f$ is the infall rate of matter onto the Galactic disk per unit area, $Z_f$ is
the abundance of infalling gas, $t$ is time, $R$ is galactocentric distance. In addition to the 
corresponding equation of LF we include the effect of diffusion (in the last term in the right hand
side of Eq (1), $D$ is the coefficient of heavy elements diffusion). The matter is that, due to
stellar winds, supernovae explosions, Galactic fountains, etc., the interstellar gas is highly turbulent.
This leads to gas mixing, which tends to smooth out the abundance gradients. For a phenomenological
description of this process we introduced the diffusion term into Eq. (1).

From previous studies it is known that the influence of the SFR radial law happens to be very important
on the Galactic abundance structure. Numerous approximation formulas were proposed for it. As a rule
the  SFR $\psi$ is assumed to be proportional to some power of gas surface density $\mu_g$. However
this cannot be recognized as completely satisfactory. Indeed, it is well known that supernovae of all
types - sources of heavy elements - are strongly concentrated in spiral arms (Bartunov et al, 1994).
Hence synthesis of heavy elements in a given volume of matter takes only place when the volume occurs
to be inside an arm. To take into account the effects of spiral arms and corotation on chemical
enrichment we represent the SFR in a manner similar to Oort (1974), Wyse \& Silk (1989) and PC:
\begin{equation}
\psi = \beta \mu_g^k |\Omega(R)-\Omega_P|\Theta(\nu)
\end{equation}
where  $\Omega(R)$ is the angular rotation velocity of the Galactic disk, $\Omega_P$ is the rotation
 velocity of the Galactic density waves responsible for spiral arms (recall that the spiral pattern 
rotates as a solid body, i.e. $\Omega_P = const$, whereas the Galactic disk rotates differentially,
 i.e. $\Omega$ is a function of $R$, the distance $R_C$ where both the velocities coincide, i.e.
$\Omega(R_C) = \Omega_P$, is called the {\it corotation radius}), $\beta$ and $k$ are some constants
 (we select the product $P_Z \beta$  so that $Z(R_{\odot})=Z_{\odot},
$\footnote{According to various authors (see e.g. PC and references therein) the solar metallicity
 is greater than that of its surroundings. However Andrievsky et al. (2002a) showed that the Sun has
 a  metallicity which is typical of its vicinity.}
 and consider two cases $k = 1$ and $k = 1.4$  Kennicutt, 1998), $\Theta$ is a cutoff factor (see below).
 
The factor $|\Omega - \Omega_P|$ in Eq (2) means a frequency of crossing spiral arms by some element of
matter of Galactic disk. It appears due to the fact that the enrichment by heavy elements takes only 
place when the element of Galactic matter occurs to be within a spiral arms. An equivalent interpretation
is that the SFR depends on the rate at which the gas penetrates the spiral arms.

Wyse \& Silk (1989) and PC considered the case when the corotation resonance is situated at the very
end of the Galaxy (model of Lin et al, 1969). So, they simply neglected by the value $\Omega_P$ in Eq (2). 

In our paper we do not neglect  $\Omega_P$ and consider several cases: with the corotation close to the
position of the Sun ($R_C \approx R_{\odot}=8.5 \, kpc$); corotation near the galactic center ($R_C = 4 \,
 kpc$) and far corotation ($R_C = 14 \, kpc$).

The cutoff factor $\Theta$ is required because the self-sustained Galactic density waves are known
 to exist only in the so-called wave zone between inner Lindblad resonance $R_{in}$ where $\nu(R_{in})=
  -1$ and the outer Lindblad resonance $R_{out}$ where $\nu(R_{out})= +1$ (Lin et al., 1969). So,
 we define $\Theta = 1$ for $|\nu| < 1$ and $\Theta = 0$ for $|\nu| > 1$. Here,
$\nu = m(\Omega_P - \Omega)/\kappa$ is the dimensionless wave frequency, $m$ is the number of arms,
   $\kappa=2\Omega(1+Rd\Omega/dR/2\Omega)^{1/2}$ is the epicyclic frequency.
   
The expression (2) for SFR means that we only consider an ordered star formation process triggered 
by spiral arms. Besides this, in the Galaxy, some sporadic (or spontaneous) star formation may take place.
Hence, in general $\psi$ should include the corresponding term. But we restrict ourselves to the 
representation (2) for the SFR; in other words, we consider that the spontaneous SFR is negligible 
compared to that induced by the spiral arms.

From the above equation, for a given set of input functions, we can compute the time history of chemical
 evolution of the Galactic disk. However, in this paper we only consider the simplest asymptotic solution. 
 Indeed, due to diffusion term in Eq (1), we expect that $\partial Z/ \partial t \to 0$ for $t \to \infty$. 
 A typical diffusion time for the achievement of the asymptotic state is $t_D \sim L^2/D$
 where $L$ is a typical length, say $L \sim 1 \, kpc$. Let us estimate $t_D$ and show that we may use
 the asymptotic regime.  For our phenomenological theory we consider  interstellar gas clouds as particles.
 So, we use the gas kinetic theory for the estimation of the diffusion coefficient: 
$$\alpha \equiv  \mu_g D = \frac{1}{3\pi\sqrt 2} \frac{m_C}{a^2}H V_T$$
where $m_C$ is the mass of a cloud, $a$ is its typical radius, $H$ is the thickness of gas layer ($H = 0.13 \,
 kpc$) and $V_T$ is the thermal velocity of a cloud ($V_T = 6.6 \,km\,s^{-1}$, Stark, 1984). Following 
 Falgarone \& Puget (1986; see also Elmegreen, 1987) we adopt $m_C/a^2 = 100 \, M_{\odot}/pc^2$. Notice 
 that for constant $H$ and $V_T$ the value $\alpha$ is constant as well. For a typical $\mu_g \sim 10 \,
  M_{\odot}/pc^2$ $t_D \sim 1.6 \, Gyr$. So, in the major part of the Galaxy we may use the above asymptotic
 approximation although in a more refined theory departures from it should be investigated.

As boundary conditions we adopt: at the "end" of the galactic disk, say at $R = R_G = 25 \, kpc$ $Z=Z_f$
($Z_f = 0.17 \, Z_{\odot}$, $Z_{\odot}=0.02$ Tinsley, 1980)\footnote {Results depend weakly on the exact value $R_G$ 
in the range 20 - 30 kpc} and at the galactic center the abundance must be finite, therefore 
$\partial Z/ \partial R = 0$ for $R = 0$ (in what follows we shall not be interested in the abundance
 structure of the central part of the Galaxy since chemical composition there is masked by very complicated
 processes due to the bar,  the bulge etc.).

Let us now discuss the functions entering Eq (1). For the infall rate we adopt: 
$$f = f_{\odot} exp(-\frac{R-R_{\odot}}{\Delta}),$$
where $f_{\odot} = 3.5 \, M_{\odot}pc^{-2}Gyr^{-1}$ (Haywood et al, 1997). 

For the surface gas density we use a simple representation 
$$\mu_g = \mu_0 exp(-\frac{R}{d}-\frac{d}{R})$$
which imitates the hole in the central part of the disk with maximum $\mu_g$ at $R = d$ and is close
 to observed radial distribution of interstellar gas (Dame, 1993) for $d = 4 \, kpc$, the dependence
being normalized as to give $\mu_g(R_{\odot})= 10 \, M_{\odot} pc^{-2}$ (Haywood et al, 1997). 

In all computations the rotation curve of Allen \& Santillan (1994) with $R_{\odot} = 8.5 \,
 kpc$ and $V_{rot}(R_{\odot}) = 220 \, km\,s^{-1}$ was used. 


\section{Results and Discussion}

First of all we consider a model with corotation close to the Sun ($R_C = R_{\odot})$ and radial inflow
 $u = 0$. Our experiments with $\Delta$ in the range 4 - 5 kpc (close to usually adopted) give similar
 results. In Fig 1 by solid lines are shown the abundance patterns for $k = 1.4$ (thick line) and $ k = 1$
 (thin line) for $\Delta = 5 \, kpc$. To feel the effects of corotation, diffusion and cuttoff $\Theta$-term 
on chemical enrichment we give the stationary solution of Eq (1) neglecting the diffusion ($k = 1.4$ - dotted
line; $k = 1$ - dashed line). Note abrupt decrease of abundance in the last two cases beyond the Lindblad
resonances due to cuttoff $\Theta$-term and the gap in the vicinity of the corotation radius. As it
was expected those sharp features are smoothed out due to diffusion.

In Fig 2 thick line is the same as in Fig 1 but the dashed line represent the solution of Eq (1) for the 
corotation slightly closer to the galactic center, namely $R_C = 8 \, kpc$ and $k = 1.4$. Both the curves
are superimposed on the observed abundance derived from Cepheids by Andrievsky et al. (2002 b). From 
this Figure it can be seen that the model with the corotation close to the Sun describes rather well
the bimodal abundance gradient derived by Andrievsky et al (2002 b), with a plateau beyond the solar
circle and a steep gradient inside of it.

In Fig 3 is shown the influence of radial gas flow: i) with constant velocity $u$: $u = + 0.5 \, km s^{-1}$
and $u = - 0.5 \, km s^{-1}$ and ii) with outflow from the corotation with the same amplitude $0.5 \, km s^{-1}$ (L\'epine et al, 2001).
Thick line is the same as in Fig 1.

From the above Fig 1-3 one can see that it is possible to explain the abundance features like bimodal
abundance structure of Andrievsky et al (2002) by means of combined influence of spiral arms, closeness
of corotation to the solar galactocentric distance and radial gas flow on chemical enrichment.

In Figs 4,5 by solid lines the results are given for models with the corotation near the galactic center
 ($R_C = 4 \, kpc$, Fig 4) and far corotation ($R_C = 14 \, kpc$, Fig 5). In both pictures dotted
 lines represent stationary solutions obtained by neglecting the diffusion term.  From these picture it is obvious that
 these models for Galactic spiral waves are not consistent with the observed abundance pattern.

\acknowledgments

We are grateful to S.M.Andrievsky for very useful discussion of the problem.


\clearpage

Figure captions

Fig. 1 - The theoretical radial abundance distribution for the corotation radius close to the Sun position, i.e. $R_C = R_{\odot} = 8.5 \, kpc$. Solid lines: thick - $k = 1.4$, thin - $k = 1$. Dotted and dashed lines describe the stationary solution of Eq (1) neglecting the diffusion term for $k = 1.4$ and $k = 1$ correspondingly. Radial inflow $u = 0$. 


Fig. 2 - The theoretical abundance gradient (solid and dashed lines) superimposed on the observational data derived by Andrievsky et al (2002 b) for Cepheids (crosses). The thick line is the same as in Fig 1, the dashed one corresponds to the corotation radius slightly closer to the Galactic center then in the previous case, namely $R_C = 8 \, kpc$. Radial inflow $u = 0$. 

Fig. 3 - The theoretical abundance gradient taking into account the radial inflow: i) $u = + 0.5 \,km s^{-1}$ - dashed line; ii) $u = - 0.5 \, km s^{-1}$ - dotted line; iii) outflow from the corotation (L\'epine et al, 2001) with the amplitude $0.5 \, km s^{-1}$ - dashed - dotted line, the corotation radius being situated at the Sun galactocentric distance. The thick solid line is the same as in Fig. 1. 


Fig. 4 - The theoretical abundance gradient for the model with the corotation near the galactic center, $R_C = 4 \, kpc$ and $k=1.4$ - solid line. Dashed line represents the stationary solution neglecting the diffusion. 


Fig. 5 - The same as Fig. 4 but for the far corotation model, $R_C = 14 \, kpc$. 


\begin{references} 

\reference{AS} Allen, C., \& Santill\'an, A., 1991, Rev. Mexicana Astron. Astrof 22, 255.

\reference{AL} Amaral, L.H., \& L\'epine, J.R.D., 1997, MNRAS, 286, 885

\reference{AKLBL} Andrievsky, S.M., Kovtyukh, V.V., Luck, R.E., L\'epine, J.R.D., Bersier, D., Maciel, W.J., Barbuy, B., Klochkova, V.G., Panchuk, V.E., \% Karpischek R.U., 2002 a, A\&A 381, 32.

\reference{AKLBL} Andrievsky, S.M., Bersier, D., Kovtyukh, V.V., Luck, R.E., Maciel, W.J., L\'epine, J.R.D., \& Beletsky, Yu.V., 2002 b, A\&A 384, 140.

\reference{BTF} Bartunov, O.S., Tsvetkov, D.Yu., \&  Filimonova, I.V., 1994, Publ. Astron. Soc. Pacific 106, 1276.

\reference{CMMP} Caputo, F., Marconi, M., Musella, I., \& Pont F., 2001, A\&A 372, 544.

\reference{CM} Cr\'ez\'e, M., \& Mennessier, M.O., 1973, A\&A 27, 281.

\reference{Da} Dame, T.M., 1993, in {\it Back to the Galaxy}, eds Holt, S., \& Verter, F., p. 267.

\reference{D} Dehnen, W., 2000, AJ 119, 800.

\reference{E} Elmegreen, B., 1987, in {\it Interstellar Processes}, eds Hollenbach, D.J., \& Thronson, H.A., Jr. (Dordrecht: Reidel), p. 259

\reference{EG} Englmaier, P., \& Gerhard, O., 1999, MNRAS 304, 512.

\reference{FP} Falgarone, E., \& Puget, J.L., 1985, A\&A 162, 235.

\reference{FFT} Fern\'andez, D., Figueras, F., \& Torra, J., 2001, A\&A 372, 833.

\reference{H} Harris, H., 1981, AJ 86, 707.

\reference{HP} Harris, H., \& Pilachowski, C., 1984, ApJ 282, 655.

\reference{HRC} Haywood, M., Robin, A.C., \& Cr\'ez\'e, M., 1997, A\&A 320, 440.

\reference{K} Kennicutt, R.C., Jr, 1998, ApJ 498, 541.

\reference{LF} Lacey, C.G.,  \& Fall, S.M., 1985, ApJ 290, 154 (LF).

\reference{LV} Leitch, E.M., \& Vasisht, G., 1998, New Astron 3, 51.

\reference{LMD} L\'epine, J.R.D., Mishurov, Yu.N., \& Dedikov, S.Yu., 2001, ApJ 546, 234.

\reference{LYS} Lin, C.C., Yuan, C., \& Shu, F.H., 1969, ApJ 155, 721.

\reference{LOMLS} Linden, von S., Otmianowska-Mazur, K., Lesch, H., \& Skupniewicz, G., 1998, A\&A 333, 79.

\reference{MQ} Maciel, W.J., \& Quireza, C., 1999, A\&A 345, 629.

\reference{MMS} Marochnik, L.S., Mishurov, Yu.N., \& Suchkov, A.A., 1972, Aph\&SS 19, 285.

\reference{MF} Martinet, L., \& Friedly, D., 1997, A\&A 323, 363.

\reference{MPS} Mishurov, Yu.N., Pavlovskaya, E.D., \& Suchkov A.A., 1979, Astron. Zhurnal (Soviet) 56, 268.

\reference{MZDMR} Mishurov, Yu.N., Zenina, I.A., Dambis, A.K., Mel'nik, A.M., \& 
Rastorguev, A.S., 1997, A\&A 323, 775.

\reference{MZ} Mishurov, Yu.N., \& Zenina, I. A., 1999 a, A\&A 341, 81.

\reference{MiZe} Mishurov, Yu.N., \& Zenina, I.A., 1999 b, Astronomy Reports 43, 487.

\reference{NM} Nelson, A.H., \& Matsuda, T., 1977, MNRAS 179, 663.

\reference{O} Oort, J.H., 1974, in IAU Sympsium 58, {\it The Formation and Dynamics of Galaxies}, ed. Shakeshaft, J.R. (Dordrecht:Reidel), p. 375.

\reference{PC} Portinari, L., \& Chiosi, C., 1999, A\&A 350, 827 (PC).

\reference{SCREH} Simpson, J.P., Colgan, S.W.J., Rubin, R.H., Erickson, E.F., \& Haas, M.R., 1995, ApJ 444, 721.

\reference{SR} Smart, S.J., \& Rolleston, W.R., 1997, ApJ 481, L47.

\reference{S} Stark, A.A., 1984, ApJ 281, 624.

\reference{T} Tinsley, B.M., 1980, Fund. Cosmic Phys. 5, 287.

\reference{TAAT} Twarog, B.A., Ashman, K.M., Anthony-Twarog, B.J., 1997, AJ 114, 2556.

\reference{VE}  V\'ilchez, J.M., \& Esteban, C., 1996, MNRAS 280, 720.

\reference{W} Weinberg, M.D., 1994, ApJ 420, 597.

\reference{WS} Wyse, F.G., \& Silk, J., 1989, ApJ 339, 700.

\end{references}
\end{document}